\documentclass{article}

\usepackage[numbers,sort&compress]{natbib}

\makeatletter

\usepackage{etex}

\usepackage{zref-savepos}

\newcounter{mnote}%[page]

\def\xmarginnote{%
  \xymarginnote{\hskip -\marginparsep \hskip -\marginparwidth}}

\def\ymarginnote{%
  \xymarginnote{\hskip\columnwidth \hskip\marginparsep}}

\long\def\xymarginnote#1#2{%
\vadjust{#1%
\smash{\hbox{{%
        \hsize\marginparwidth
        \@parboxrestore
        \@marginparreset
\footnotesize #2}}}}}

\def\mnoteson{%
\gdef\mnote##1{\refstepcounter{mnote}\label{##1}%
  \zsavepos{##1}%
  \ifnum20432158>\number\zposx{##1}%
  \xmarginnote{{\color{blue}\bf $\langle$\arabic{mnote}$\rangle$}}% 
  \else
  \ymarginnote{{\color{blue}\bf $\langle$\arabic{mnote}$\rangle$}}%
  \fi%
}
  }
\gdef\mnotesoff{\gdef\mnote##1{}}
\mnoteson
\mnotesoff

\usepackage{framed}
\usepackage{subcaption}
\usepackage{comment}
\usepackage[svgnames,dvipsnames]{xcolor}
\usepackage[all]{xy}
\usepackage{tikz}
\usetikzlibrary{positioning,matrix,through,calc,arrows,fit,shapes,decorations.pathreplacing,decorations.markings,plotmarks}
\usepackage{pgfplotstable}
\usepackage{pgfplots}
\tikzstyle{block} = [draw,fill=blue!20,minimum size=2em]

\usepackage{qsymbols,amssymb,mathrsfs}
\usepackage[standard,thmmarks]{ntheorem}
\theoremstyle{plain}
\theoremsymbol{\ensuremath{_\vartriangleleft}}
\theorembodyfont{\itshape}
\theoremheaderfont{\normalfont\bfseries}
\theoremseparator{}

\theoremstyle{nonumberplain}
\theoremheaderfont{\scshape}
\theorembodyfont{\normalfont}
\theoremsymbol{\ensuremath{_\blacktriangleleft}}

\theoremnumbering{arabic}
\theoremstyle{plain}
\usepackage{latexsym}
\theoremsymbol{\ensuremath{_\Box}}
\theorembodyfont{\itshape}
\theoremheaderfont{\normalfont\bfseries}
\theoremseparator{}

\theorembodyfont{\upshape}
\theoremprework{\bigskip\hrule}
\theorempostwork{\hrule\bigskip}
\usepackage[overload]{empheq} % no \intertext and \displaybreak

\let\iftwocolumn\if@twocolumn
\g@addto@macro\@twocolumntrue{\let\iftwocolumn\if@twocolumn}
\g@addto@macro\@twocolumnfalse{\let\iftwocolumn\if@twocolumn}

\mathtoolsset{showonlyrefs=false,showmanualtags}
\let\underbrace\LaTeXunderbrace % adapt spacing to font sizes
\let\overbrace\LaTeXoverbrace
\renewcommand{\eqref}[1]{\textup{(\refeq{#1})}} % eqref was not allowed in
\newtagform{brackets}[\textbf]{(}{)}   % new tags for equations
\usetagform{brackets}
\usepackage[Smaller]{cancel}

\usepackage{url}
\usepackage{graphicx,psfrag}
\graphicspath{{figure/}{image/}} % Search path of figures

\usepackage{diagbox} % \backslashbox{}{} for slashed entries
\usepackage{algorithm2e}
\usepackage{listings} 
\lstdefinelanguage{Maple}{
  morekeywords={proc,module,end, for,from,to,by,while,in,do,od
    ,if,elif,else,then,fi ,use,try,catch,finally}, sensitive,
  morecomment=[l]\#,
  morestring=[b]",morestring=[b]`}[keywords,comments,strings]
\DeclareMathAlphabet{\mathpzc}{OT1}{pzc}{m}{it}
\usepackage{upgreek} % \upalpha,\upbeta, ...
\usepackage{dsfont}  % \mathds

\def\multi@nostar#1#2{%
  \expandafter\def\csname multi#1\endcsname##1{%
    \if ##1.\let\next=\relax \else
    \def\next{\csname multi#1\endcsname}     
    \expandafter\newcommand\csname #1##1\endcsname{#2}
    \fi\next}}

\def\multi@star#1#2{%
  \expandafter\def\csname #1\endcsname##1{#2}
  \multi@nostar{#1}{#2}
}

\newcommand{\multi}{%
  \@ifstar \multi@star \multi@nostar}

\multi*{rm}{\mathrm{#1}}
\multi*{mc}{\mathcal{#1}}
\multi*{op}{\mathop {\operator@font #1}}
\multi*{ds}{\mathds{#1}}
\multi*{set}{\mathcal{#1}}
\multi*{rsfs}{\mathscr{#1}}
\multi*{pz}{\mathpzc{#1}}
\multi*{M}{\boldsymbol{#1}}
\multi*{R}{\mathsf{#1}}
\multi*{RM}{\M{\R{#1}}}
\multi*{bb}{\mathbb{#1}}
\multi*{td}{\tilde{#1}}
\multi*{tR}{\tilde{\mathsf{#1}}}
\multi*{trM}{\tilde{\M{\R{#1}}}}
\multi*{tset}{\tilde{\mathcal{#1}}}
\multi*{tM}{\tilde{\M{#1}}}
\multi*{baM}{\bar{\M{#1}}}
\multi*{ol}{\overline{#1}}

\multirm  ABCDEFGHIJKLMNOPQRSTUVWXYZabcdefghijklmnopqrstuvwxyz.
\multiol  ABCDEFGHIJKLMNOPQRSTUVWXYZabcdefghijklmnopqrstuvwxyz.
\multitR   ABCDEFGHIJKLMNOPQRSTUVWXYZabcdefghijklmnopqrstuvwxyz.
\multitd   ABCDEFGHIJKLMNOPQRSTUVWXYZabcdefghijklmnopqrstuvwxyz.
\multitset ABCDEFGHIJKLMNOPQRSTUVWXYZabcdefghijklmnopqrstuvwxyz.
\multitM   ABCDEFGHIJKLMNOPQRSTUVWXYZabcdefghijklmnopqrstuvwxyz.
\multibaM   ABCDEFGHIJKLMNOPQRSTUVWXYZabcdefghijklmnopqrstuvwxyz.
\multitrM   ABCDEFGHIJKLMNOPQRSTUVWXYZabcdefghijklmnopqrstuvwxyz.
\multimc   ABCDEFGHIJKLMNOPQRSTUVWXYZabcdefghijklmnopqrstuvwxyz.
\multiop   ABCDEFGHIJKLMNOPQRSTUVWXYZabcdefghijklmnopqrstuvwxyz.
\multids   ABCDEFGHIJKLMNOPQRSTUVWXYZabcdefghijklmnopqrstuvwxyz.
\multiset  ABCDEFGHIJKLMNOPQRSTUVWXYZabcdefghijklmnopqrstuvwxyz.
\multirsfs ABCDEFGHIJKLMNOPQRSTUVWXYZabcdefghijklmnopqrstuvwxyz.
\multipz   ABCDEFGHIJKLMNOPQRSTUVWXYZabcdefghijklmnopqrstuvwxyz.
\multiM    ABCDEFGHIJKLMNOPQRSTUVWXYZabcdefghijklmnopqrstuvwxyz.
\multiR    ABCDEFGHIJKL NO QRSTUVWXYZabcd fghijklmnopqrstuvwxyz.
\multibb   ABCDEFGHIJKLMNOPQRSTUVWXYZabcdefghijklmnopqrstuvwxyz.
\multiRM   ABCDEFGHIJKLMNOPQRSTUVWXYZabcdefghijklmnopqrstuvwxyz.

\newcommand{\dotleq}{\buildrel \textstyle  .\over {\smash{\lower
      .2ex\hbox{\ensuremath\leqslant}}\vphantom{=}}}
\newcommand{\dotgeq}{\buildrel \textstyle  .\over {\smash{\lower
      .2ex\hbox{\ensuremath\geqslant}}\vphantom{=}}}

\newcommand{\bM}{\begin{bmatrix}}
\newcommand{\eM}{\end{bmatrix}}
\newcommand{\bSM}{\left[\begin{smallmatrix}}
\newcommand{\eSM}{\end{smallmatrix}\right]}
\renewcommand*\env@matrix[1][*\c@MaxMatrixCols c]{%
  \hskip -\arraycolsep
  \let\@ifnextchar\new@ifnextchar
  \array{#1}}

\newqsymbol{`N}{\mathbb{N}}
\newqsymbol{`R}{\mathbb{R}}
\newqsymbol{`Z}{\mathbb{Z}}

\newqsymbol{`|}{\mid}
\newqsymbol{`8}{\infty}
\newqsymbol{`1}{\left}
\newqsymbol{`2}{\right}
\newqsymbol{`6}{\partial}
\newqsymbol{`0}{\emptyset}
\newqsymbol{`-}{\leftrightarrow}
\DeclarePairedDelimiter\Set{\{}{\}}
\newcommand{\imod}[1]{\allowbreak\mkern10mu({\operator@font mod}\,\,#1)}

\newcommand{\threecols}[3]{
\hbox to \textwidth{%
      \normalfont\rlap{\parbox[b]{\textwidth}{\raggedright#1\strut}}%
        \hss\parbox[b]{\textwidth}{\centering#2\strut}\hss
        \llap{\parbox[b]{\textwidth}{\raggedleft#3\strut}}%
    }% hbox 
}

\newcommand{\reason}[2][\relax]{
  \ifthenelse{\equal{#1}{\relax}}{
    \left(\text{#2}\right)
  }{
    \left(\parbox{#1}{\raggedright #2}\right)
  }
}

\let\SavedDoubleVert\relax
\let\protect\relax
{\catcode`\|=\active
  \xdef\extendvert{\protect\expandafter\noexpand\csname extendvert \endcsname}
  \expandafter\gdef\csname extendvert \endcsname#1{\mskip-5mu \left.%
      \ifx\SavedDoubleVert\relax \let\SavedDoubleVert\|\fi
     \:{\let\|\SetDoubleVert
       \mathcode`\|32768\let|\SetVert
     #1}\:\right.\mskip-5mu}
}
\def\SetVert{\@ifnextchar|{\|\@gobble}% turn || into \|
    {\egroup\;\mid@vertical\;\bgroup}}
\def\SetDoubleVert{\egroup\;\mid@dblvertical\;\bgroup}

\begingroup
 \edef\@tempa{\meaning\middle}
 \edef\@tempb{\string\middle}
\expandafter \endgroup \ifx\@tempa\@tempb
 \def\mid@vertical{\middle|}
 \def\mid@dblvertical{\middle\SavedDoubleVert}
\else
 \def\mid@vertical{\mskip1mu\vrule\mskip1mu}
 \def\mid@dblvertical{\mskip1mu\vrule\mskip2.5mu\vrule\mskip1mu}
\fi

\makeatother

\usepackage{ctable}
\usepackage{fouridx}
\usepackage{framed}
\usetikzlibrary{positioning,matrix}

\usepackage{paralist}
\usepackage{enumerate}
\usepackage{enumitem}
\setlist[1]{leftmargin=2em,itemsep=0em}

\usepackage[normalem]{ulem}

{\endMakeFramed}

\newenvironment{ybox}{
	\setlength{\FrameSep}{1.5mm}
	\setlength{\FrameRule}{0mm}
  \MakeFramed {\FrameRestore}}%
{\endMakeFramed}

\newenvironment{gbox}{
  \MakeFramed {\FrameRestore}}%
{\endMakeFramed}

{\endMakeFramed}

 {\endMakeFramed}

\DontPrintSemicolon
\SetAlgoLined
\SetKwRepeat{Struct}{struct \{}{\}}%

\SetKwArray{Parent}{parent}
\SetKwArray{Children}{children}
\SetKwArray{Node}{node}
\SetKwArray{Weight}{weight}
\SetKwArray{Rank}{rank}
\SetKwArray{Edges}{edges}
\SetKwArray{EdgesAt}{edges\_at}
\SetKwFunction{getClustersOfSize}{getClustersOfSize}
\SetKwFunction{getCriticalValues}{getCriticalValues}
\SetKwFunction{getPartition}{getPartition}
\SetKwFunction{similarity}{similarity}
\SetKwFunction{find}{find}
\SetKwFunction{merge}{merge}
\SetKwFunction{Agglomerate}{Agglomerate}
\SetKwFunction{Segregate}{Segregate}
\SetKwFunction{MMI}{MMI}
\SetKwFunction{DT}{DT}
\SetKwFunction{SFM}{SFM}
\SetKwFunction{IMMI}{IMMI}
\SetKwFunction{DMMI}{DMMI}
\SetKwFunction{MUI}{MUI}
\SetKwFunction{Redundancy}{Redundancy}
\SetKwFunction{ISplit}{ISplit}
\SetKwFunction{DSplit}{DSplit}
\SetKwProg{myproc}{procedure}{:}{end}
\SetKwProg{myfn}{function}{:}{end}

\SetCommentSty{mycommfont}

     \usepackage[preprint]{neurips_2019}

\usepackage[utf8]{inputenc} % allow utf-8 input
\usepackage[T1]{fontenc}    % use 8-bit T1 fonts
\usepackage{hyperref}       % hyperlinks
\usepackage{url}            % simple URL typesetting
\usepackage{booktabs}       % professional-quality tables
\usepackage{amsfonts}       % blackboard math symbols
\usepackage{nicefrac}       % compact symbols for 1/2, etc.
\usepackage{microtype}      % microtypography

\usepackage{pgfplots}

\title{Neural Entropic Estimation: \\ A faster path to mutual information estimation}

\author{
Chung Chan\thanks{This work is supported by a grant from the University Grants Committee of the Hong Kong Special Administrative Region, China (Project No. 21203318).}\\
Department of Computer Science\\
City University of Hong Kong\\
\texttt{chung.chan@cityu.edu.hk}\\
\And 
Ali Al-Bashabsheh\\
Big Data and Brain Computing\\
Beihang University\\
\And
Hing Pang Huang\\
Department of Computer Science\\
City University of Hong Kong\\
\And
Michael Lim\\
Department of Computer Science\\
City University of Hong Kong\\
\And
Da Sun Handason Tam\\
Department of Computer Science\\
City University of Hong Kong\\
\And
Chao Zhao\\
Department of Computer Science\\
City University of Hong Kong\\
}

\begin{document}

\maketitle

\begin{abstract}
We point out a limitation of the mutual information neural estimation (MINE) where the network fails to learn at the initial training phase, leading to slow convergence in the number of training iterations. To solve this problem, we propose a faster method called the mutual information neural entropic estimation (MI-NEE). Our solution first generalizes MINE to estimate the entropy using a custom reference distribution. The entropy estimate can then be used to estimate the mutual information. We argue that the seemingly redundant intermediate step of entropy estimation allows one to improve the convergence by an appropriate reference distribution. In particular, we show that MI-NEE reduces to MINE in the special case when the reference distribution is the product of marginal distributions, but faster convergence is possible by choosing the uniform distribution as the reference distribution instead. Compared to the product of marginals, the uniform distribution introduces more samples in low-density regions and fewer samples in high-density regions, which appear to lead to an overall larger gradient for faster convergence.

\end{abstract}

\section{Introduction}
\label{sec:introduction}

The measure of mutual information~\cite{shannon48} has significant applications in data mining~\cite{guyon2003introduction, estevez2009normalized}. An advantage of mutual information over other distances or similarity measures is that, in addition to linear correlation,
%, not only captures linear correlation between two feature vectors but also 
it also captures non-linear functional or statistical dependency between different features. Therefore, it has been used to select, extract and cluster features~\cite{peng05,hjelm2018learning} in an unsupervised way. The measure has firm theoretic ground in information theory, and can be understood as the fundamental limits of the rate-distortion function~\cite{shannon59coding}, channel capacity~\cite{shannon48}, and secrecy capacity~\cite{ahlswede93}. 

To apply mutual information to practical scenarios in data mining, one has to estimate it from data samples with limited or no knowledge of the underlying distribution. Mutual information estimation is a well-known difficult problem, especially when the feature vectors are continuous or in a high dimensional space~\cite{ chow2005estimating, peng05}. Despite the limitation of the well-known histogram approach~\cite{steuer2002mutual,paninski2003estimation}, there are various other estimation methods, including different density estimations using a kernel~\cite{moon1995estimation} and the nearest-neighbor distance~\cite{kraskov04}. 

A more recent work considers iterative estimation using a neural network, called the mutual information neural estimation (MINE)~\cite{belghazi18}. Compared to other approaches, MINE appears to inherit the generalization capability of neural network and can work without careful choice of parameters. However, as the neural network needs to be trained iteratively by a gradient descent algorithm, one has to monitor the convergence of the estimate and decide when to stop. If the convergence rate is slow, one may have to wait for a long time and terminate prematurely, which can result in underfitting. 
%Indeed, we discovered simple bivariate distributions where MINE converged very slowly. The objective of this work is to understand and resolve this obvious short-coming of MINE in the bivariate case, which is essential before applying the neural estimation to higher dimensional cases. In particular, we propose an alternative route of neural estimation that drastically improves the convergence rate. Roughly speaking, MINE uses a neural network to estimate the divergence from the joint distribution to the product of marginal distributions. If we replace the product of the marginal distributions by a known uniform reference distribution, we can obtain an estimate of the joint entropy instead of the mutual information, but the convergence rate turns out to be much faster. Since the mutual information can be computed simply from the joint and marginal entropies, and the marginal entropy can be estimated more easily than the joint, we can obtain a faster mutual information estimate than MINE. 
Indeed, we discovered a simple bivariate mixed gaussian distribution where MINE converged very slowly, and the problem is more serious in the higher dimensional cases. The objective of this work is to understand and resolve this short-coming, which is essential before applying the neural estimation to real-world datasets that often have very high dimensions.  Despite the huge success in the use of neural networks for various machine learning applications~\cite{krizhevsky2012imagenet,goodfellow2014generative,silver2016mastering,peters2018deep,pouyanfar2018survey}, the current understanding of neural network is limited. A proof of the generalization capability is known only for a very simple model~\cite{brutzkus2017sgd}. % Other published limitations of MINE
%"Entropy and mutual information in models of deep neural networks" (NIPS2018) says MINE is computationally hard (only mentioned).

We propose an alternative route of neural estimation, called the mutual information neural entropic estimation (MI-NEE), that drastically improves the convergence rate. Roughly speaking, MINE uses a neural network to estimate the divergence from the joint distribution to the product of marginal distributions. If we replace the product of the marginal distributions by a known uniform reference distribution, we can obtain an estimate of the joint entropy instead of the mutual information, but the convergence rate turns out to be much faster. Since the mutual information can be computed simply from the joint and marginal entropies, and the marginal entropies can be estimated more easily than the joint entropy, we can obtain a faster mutual information estimate than MINE. 

Our approach, in the use of a custom reference distribution, may resemble contrastive / ratio estimation methods \cite[Sec. 12.2.4, pp~495--497]{hastie2009elements}, \cite{gutmann2012noise}, which provides a neural estimation of the ratio between the unknown and the reference distributions (often by casting the unsupervised problem as a classification problem). However, the objective here is to estimate the KL divergence between the unknown distribution and the reference distribution by maximizing a lower bound, namely, the KL divergence between the neural network's parameterized distribution and the reference distribution. For more details on the contrastive approach and its relation to MINE, see \cite{oord2018representation, poole2018variational}. Other neural network approaches to estimating density with respect to a reference distribution exist, e.g., in \cite{dinh2014nice, dinh2016density}, a neural network is used to obtain a deterministic map between a latent random variable with a known distribution and the data.

Detailed derivations of our approach will be given in Section~\ref{sec:solution} following the problem formulation in~Section~\ref{sec:problem}. Some experimental results will be given in Section~\ref{sec:experiment}.

\section{Problem formulation}
\label{sec:problem}

We use a sans serif capital letter $\RZ$ to denote a random vector/variable and the same character $Z$ in the normal math font for its alphabet set.  $p_{\RZ}$ denotes the distribution of $\RZ$, which is a pdf if $\RZ$ is continuous. The support $\op{supp}(p_{\RZ})$ of (the distribution of) $\RZ$ is the subset of values in $Z$ with strictly positive probability density. $E$ denotes the expectation operation. For simplicity, all the logarithms are natural logarithms, and so information quantities such as entropy and mutual information are measured in nats instead of bits.

\paragraph{Mutual information estimation} Given continuous random vectors/variables $\RX$ and $\RY$ with unknown pdf $p_{\RX\RY}(x,y)$ for $(x,y)\in X \times Y$, the goal is to estimate the following Shannon's mutual information from $N\geq 1$ i.i.d.\ samples $(\RX_1,\RY_1),\dots,(\RX_N,\RY_N)$ of $(\RX,\RY)$:
\begin{subequations}
\label{eq:I}
\begin{align}
    I(\RX \wedge \RY) &= D(p_{\RX\RY}\| p_{\RX}p_{\RY}) = E`1[\ln \frac{p_{\RX\RY} (\RX,\RY)}{p_{\RX}(\RX) p_{\RY}(\RY)}`2] \label{eq:I1}\\
    &= H(\RX) + H(\RY) - H(\RX,\RY), \label{eq:I2}
\end{align}
\end{subequations}
where $D$ and $H$ denote the information divergence and entropy respectively defined as
\begin{align}
    D(p_{\RZ}\|p_{\RZ'}) &:= E`1[\ln \frac{p_{\RZ}(\RZ)}{p_{\RZ'}(\RZ)} `2], \label{eq:D}\\
    H(\RZ) &:= E`1[\ln \frac1{p_{\RZ}(\RZ)}`2].\label{eq:H}
\end{align}

MINE~\cite{belghazi18} estimates $I(\RX\wedge \RY)$ by rewriting the divergence in~\eqref{eq:I1} as a maximization over a functional and uses a neural network to optimize the functional iteratively. In contrast, we estimate the mutual information by neural estimation of the entropies in \eqref{eq:I2}. 

\paragraph{Entropy estimation} Given a continuous random vector/variable $\RZ$ with unknown pdf, we want to estimate $H(\RZ)$ from $N$ i.i.d.\ samples $\RZ_1,\dots,\RZ_N$ of $\RZ$. 

With $\RZ$ chosen to be $\RX$, $\RY$, and  $(\RX,\RY)$ respectively, we obtain estimates of all the entropy terms in \eqref{eq:I2}, and therefore, the desired estimate of the mutual information.

% The idea of MINE~\eqref{} is to apply the following variational formula to the divergence in \eqref{eq:I1}:
% \begin{Proposition}
% For any random variables $\RZ$ and $\RZ'$ where $Z\subseteq Z'$ and
% \begin{align*}
%     p_{\RZ'}(z) &>0 \forall z\in Z: p_{\RZ}(z)>0,
% \end{align*}
% we have 
% \begin{align}
%     D(p_{\RZ}\|p_{\RZ'}) = \sup_{f:Z\mapsto `R} E`1[f(\RZ)`2] - \ln E`1[ e^{f(\RZ')}`2].
% \end{align}
% Furthermore, $f$ is optimal if and only if
% \begin{align}
%     f(z) = \ln \frac{p_{\RZ}(z)}{p_{\RZ'}(z)} + c \quad \forall z\in Z\label{eq:f*}
% \end{align}
% for some constant $c$.
% \end{Proposition}

\section{Neural entropic estimation}
\label{sec:solution}

%We derive the neural estimation~\eqref{eq:est:H} of the entropy with a custom reference distribution. The desired mutual information estimation~\eqref{eq:est:I} follows a similar derivation with the uniform reference~\eqref{eq:uniform}, where MINE is argued to be a special case when the reference distribution is the product of marginal distributions. 

We derive a neural estimation of the entropy using a custom reference distribution. The desired mutual information estimation then follows from~\eqref{eq:I2}, where MINE is argued to be a special case when the reference distribution is the product of marginal distributions. We end the section by discussing estimations using the uniform reference distribution.

\subsection{Entropy estimation}
\label{sec:entropy}

To estimate the entropy of $\RZ$ using a neural network, we rewrite the entropy in terms of the divergence between $p_{\RZ}$ and a custom reference distribution $p_{\RZ'}$ as:
\begin{align}
    H(\RZ) = \underbrace{E`1[ \ln \frac1{p_{\RZ'}(\RZ)} `2]}_{`(1)} - D(p_{\RZ}\|p_{\RZ'}). \label{eq:cross}
\end{align}
%where $p_{\RZ'}$ is the custom reference distribution.
Note that the first term (the cross entropy term) in \eqref{eq:cross} can be estimated using sample average
\begin{align}
    `(1)\approx \frac1N \sum_{i=1}^N \ln \frac1{p_{\RZ'}(\RZ_i)},\label{eq:est:cross}
\end{align}
which is an unbiased estimate because $p_{\RZ'}$ is a known pdf. For the formula to be valid, the divergence should be bounded, which requires
\begin{align}
  \op{supp}(p_{\RZ})\subseteq \op{supp}(p_{\RZ'}).\label{eq:supp}
\end{align}
Other than the above restriction, however, one is free to choose any reference $\RZ'$ in the calculation of the entropy in \eqref{eq:cross}. Indeed, not only there is no requirement for $p_{\RZ'}$ to be close to $p_{\RZ}$, we will argue that there is a benefit in choosing $p_{\RZ'}$ to be different from $p_{\RZ}$, namely, that it can lead to a faster convergence for the neural estimate of the divergence. 

As in MINE~\cite{belghazi18}, to apply a neural network to estimate the entropy, we rewrite the divergence using the variational formula~\cite{donsker1983asymptotic} as follows:
%\begin{align}
%  \begin{split}
%    D(p_{\RZ}\|p_{\RZ'}) 
%    &\utag{\text{i}}= D(p_{\RZ}\|p_{\RZ'}) - \inf_{p_{\hat{\RZ}}\in \rsfsP(Z)} D(p_{\RZ}\|p_{\hat{\RZ}})\\
%    &\utag{ii}= \sup_{p_{\hat{\RZ}}\in \rsfsP(Z)} E`1[ \ln \frac{p_{\hat{\RZ}}(\RZ)}{p_{\RZ'}(\RZ)}`2]\\
%    &\utag{iii}= \sup_{f:Z\mapsto `R} \underbrace{`1\{E`1[f(\RZ)`2] - \ln E`1[ e^{f(\RZ')}`2]`2\}}_{`(2)}.
%  \end{split}
%  \label{eq:var:D}
%  \end{align}
\begin{subequations}
  \label{eq:var:D}
  \begin{align}
    \label{eq:var:Da}
    D(p_{\RZ}\|p_{\RZ'}) 
    &= D(p_{\RZ}\|p_{\RZ'}) - \inf_{p_{\hat{\RZ}}\in \rsfsP(Z)} D(p_{\RZ}\|p_{\hat{\RZ}})\\
    \label{eq:var:Db}
    &= \sup_{p_{\hat{\RZ}}\in \rsfsP(Z)} E`1[ \ln \frac{p_{\hat{\RZ}}(\RZ)}{p_{\RZ'}(\RZ)}`2]\\
    \label{eq:var:Dc}
    &= \sup_{f:Z\mapsto `R} \underbrace{`1\{E`1[f(\RZ)`2] - \ln E`1[ e^{f(\RZ')}`2]`2\}}_{`(2)}.
  \end{align}
\end{subequations}
In the first equality~\eqref{eq:var:Da}, the infimum is over the choices of a distribution $p_{\hat{\RZ}}$ for a random variable $\hat{\RZ}$ with alphabet set $Z$. Equality holds because the divergence $D(p_{\RZ}\|p_{\hat{\RZ}})$ is non-negative and equal to $0$ if and only if $p_{\hat{\RZ}}=p_{\RZ}$. The seemingly redundant infimum term plays an important role in the neural estimation. As can be seen in the equality~\eqref{eq:var:Db}, the term $\ln p_{\RZ}$ involving the unknown distribution $p_{\RZ}$ no longer appears inside the expectation. Instead, we have the term $\ln p_{\hat{\RZ}}$ which will be evaluated and optimized by a neural network. More precisely, suppose the neural network computes the function $f:Z\mapsto `R$, it can be turned into a probability distribution by the formula
\begin{align*}
 p_{\hat{\RZ}}(z) := \frac{p_{\RZ'}(z) e^{f(z)}}{E`1[e^{f(\RZ')}`2]} \quad \forall z\in Z,
\end{align*}
which is non-negative and integrates over $z\in Z$ to $1$. Applying this formula to the supremum in \eqref{eq:var:Db} gives the last equality~\eqref{eq:var:Dc}. Since the supremum is achieved uniquely by $P_{\hat{\RZ}}=P_{\RZ}$, it follows that $f$ is optimal to \eqref{eq:var:Dc} if and only if
\begin{align}
    f(z) = \ln \frac{p_{\RZ}(z)}{p_{\RZ'}(z)} + c \quad \forall z\in Z\label{eq:f*}
\end{align}
for some constant $c$. In summary, we have
\begin{Proposition}
  For any continuous random vector/variable $\RZ$, and any other random vector/variable $\RZ'$ with a larger support~\eqref{eq:supp}, we have
  \begin{align}
    H(\RZ) &= E`1[\ln \frac1{p_{\RZ'}(\RZ)}`2]-\sup_{f:Z\to `R} `1\{E[f(\RZ)]-\ln E`1[e^{f(\RZ')}`2]`2\}.\label{eq:H2}
  \end{align}
  Furthermore, any optimal solution $f$ must satisfy the optimality condition~\eqref{eq:f*}.
\end{Proposition}

Note that the objective function in \eqref{eq:var:Dc} can be estimated from the samples as
\begin{align}
    `(2)\approx \frac1N \sum_{i=1}^N f(\RZ_i) - \ln \frac1{N'} \sum_{i=1}^{N'} e^{f(\RZ'_i)}, \label{eq:est:D}
\end{align}
where $\RZ_1,\dots,\RZ_{N}$ are i.i.d.\ samples of $\RZ$ and $\RZ'_1,\dots,\RZ'_{N'}$ are i.i.d.\ samples of $\RZ'$. Although the estimate may have bias from the estimate of the log expectation term $\ln E`1[ e^{f(\RZ')}`2]$, we can reduce such bias by choosing $N'$ sufficiently large, which is possible since $p_{\RZ'}$ is a known pdf. 
%We will explain in the following section that 
MINE also has a similar log expectation term but the bias in the estimation of the term and its corresponding gradient may be non-negligible, as the expectation there is with respect to an unknown pdf, namely, the product $p_{\RX}p_{\RY}$ of the marginal distributions. (We will briefly revisit this in Section~\ref{subsec:unif}.)

To estimate the supremum in \eqref{eq:var:Dc}, we apply a neural network as in MINE with parameters $`q$ that outputs
\begin{align}
 f(z):=`f_z(`q) \quad \forall z\in Z.\label{eq:phi}
\end{align}
Define the loss function as the negation of $`(2)$,
\begin{align}
    L(`q):= -E`1[`f_{\RZ}(`q)`2] + \ln E`1[ e^{`f_{\RZ'}(`q)}`2]. \label{eq:L}
\end{align}
We can iteratively optimize $f$ to maximize $`(2)$ by updating $`q$ with standard gradient descent algorithms that use minibatch estimates of the gradient 
\begin{align}
 \nabla L(`q)=-E`1[\nabla`f_{\RZ}(`q)`2] + \frac{E`1[ e^{`f_{\RZ'}(`q)}\nabla`f_{\RZ'}(`q) `2]}{E`1[ e^{`f_{\RZ'}(`q)}`2]}. \label{eq:L_grad}
\end{align}
Again, the expectations in the second term can be estimated by any number of samples from the known reference distribution $p_{\RZ'}$, and so the bias from the estimate of the expectation in the denominator can be made negligible if desired. In practice, the stochasticity involved in the minibatch estimates somehow avoids overfitting even with an over-parameterized neural network~\cite{zhang2016understanding, brutzkus2017sgd}, and one can often converge to a good minima using a small batch size~\cite{masters2018revisiting}. To maintain such stochasticity for large $N'$, one can simply generate new samples of $\RZ'$ for each step of the descend algorithm, which is possible as $p_{\RZ'}$ is known. 

Altogether, an estimate $H(\RZ)$ can be obtained as follows using the estimate~\eqref{eq:est:cross} of the cross entropy in \eqref{eq:cross} and the estimate~\eqref{eq:est:D} of the divergence in \eqref{eq:cross} where $f$ is optimized by training the neural network \eqref{eq:phi} for some $t\geq 0$ times using the loss function~\eqref{eq:L}. 

\paragraph{Entropy estimate} The estimate of the entropy is given by
\begin{align}
    H(\RZ)\approx \frac1N \sum_{i=1}^N \ln \frac1{p_{\RZ'}(\RZ_i)} - \frac1N \sum_{i=1}^N `f_{\RZ_i}(`q_t) + \ln \frac1{N'} \sum_{i=1}^{N'} e^{`f_{\RZ'_i}(`q_t)},\label{eq:est:H}
\end{align}
where $`q_t$ is the parameter after $t$ steps of the gradient descend algorithm. 

We remark that the above estimate is neither a lower nor an upper bound on the entropy estimate because of the possibilities of underfitting due to insufficient training and overfitting due to the use of sample estimates for the training objective. The same issue applies to MINE. Nevertheless, while one can check whether overfitting occurs using a separate validation set, it is hard to tell if there is underfitting without knowing the ground truth. Indeed, the convergence rate of the parameters $`q$ may be so slow that one may falsely think that the parameters have converged even if they have not. We found that such situation may be avoided by an appropriate choice of the reference distribution.

\subsection{Mutual information estimation}
\label{eq:I}

By expressing the mutual information in terms of the entropies in \eqref{eq:I2}, it is straightforward to obtain a mutual information estimate by estimating the entropies as explained in the previous section. We simplify the estimate further by choosing the reference distributions appropriately so that the cross entropy terms in the entropy estimates cancel out:

\begin{Proposition}
  For any continuous random vectors/variables $\RX$ and $\RY$,
  \begin{subequations}
    \label{eq:I34}
    \begin{align}
      I(\RX\wedge \RY) &= D(p_{\RX\RY}\| p_{\RX'\RY'}) - D(p_{\RX}\| p_{\RX'})-D(p_{\RY}\| p_{\RY'})
      \label{eq:I3}\\
      \begin{split}&= 
        \sup_{f_0:X\times Y\to `R} \{E`1[f_0(\RX,\RY)`2] - \ln E`1[ e^{f_0(\RX',\RY')}`2]\}\\
      &\kern1em - \sup_{f_1:X\to `R} \{E`1[f_1(\RX)`2] - \ln E`1[ e^{f_1(\RX')}`2]\}\\
      &\kern1em - \sup_{f_2: Y\to `R} \{E`1[f_2(\RY)`2] - \ln E`1[ e^{f_2(\RY')}`2]\},
      \end{split}\label{eq:I4}
    \end{align}
  \end{subequations}
  where $\RX'$ and $\RY'$ are independent random variables/vectors with larger support than $\RX$ and $\RY$, i.e., 
  %$\op{supp}(p_{\RX\RY})\subseteq \op{supp}(p_{\RX'\RY'})$.
  \begin{align}
    p_{\RX'\RY'}(x,y) = p_{\RX'}(x) p_{\RY'}(y) > 0 \quad\forall x\in X,y\in Y:p_{\RX'\RY'}(x,y)>0.\label{eq:supp:I}
  \end{align}
  Furthermore, the optimal $f_0$, $f_1$ and $f_2$ satisfy
  \begin{align}
    f_0(x,y) - f_1(x)-f_2(x) = \overbrace{\ln \frac{p_{\RX\RY}(x,y)}{p_{\RX}(x)p_{\RY}(y)}}^{i_{\RX\RY}(x,y):=} + c \label{eq:f*:I}
  \end{align}
  for some constant $c$.
\end{Proposition}

\begin{Proof}
  \eqref{eq:I3} follows from \eqref{eq:I2} and \eqref{eq:cross} with $(\RZ,\RZ')$ set to $(\RX,\RX')$, $(\RY,\RY')$, and $((\RX,\RY),(\RX',\RY'))$ respectively. Note that the cross entropy terms cancel out 
  due to the independence of $\RX'$ and $\RY'$, i.e., 
%  \begin{align}
%    p_{\RX'\RY'}(x,y) = p_{\RX'}(x)p_{\RY'}(y) \quad \forall x\in X,y\in Y.\label{eq:pX'Y'}
%  \end{align}
%  and so
$E`1[\ln \frac1{p_{\RX'\RY'}(\RX,\RY)}`2] = E`1[\frac1{\ln p_{\RX'}(\RX)}`2] + E`1[\frac1{\ln p_{\RY'}(\RY)}`2]$.
  Equation \eqref{eq:I4} follows from the variational formula~\eqref{eq:var:D}, and \eqref{eq:f*:I} follows directly from the optimality condition~\eqref{eq:f*}. 
\end{Proof}

The desired mutual information estimate can be obtained from the sample estimate of \eqref{eq:I4} with $f_0$, $f_1$, and $f_2$ optimized independently using three neural networks as described in the previous section, i.e., with the loss functions chosen as \eqref{eq:L} with $(\RZ,\RZ')$ set to $((\RX,\RY),(\RX',\RY'))$, $(\RX,\RX')$, and $(\RY,\RY')$ respectively. 

Alternatively, one can train a single neural network with three outputs, one for each $f_i$. More precisely, construct a neural network with parameters $`q$, two inputs $x\in X$ and $y\in Y$, and three outputs $`f_{x,y}:=(`f^{(0)}_{x,y},`f^{(1)}_{x},`f^{(2)}_{y})$. With
\begin{align*}
  f_i(x,y):=`f^{(i)}_{x,y}(`q) \quad \forall i\in \Set{0,1,2}, x\in X,y\in Y,
\end{align*}
we update the parameters $`q$ to minimize the sum of the loss functions \eqref{eq:L} evaluated for the three choices of $(\RZ,\RZ')$, i.e.,
\begin{align*}
  L(`q) &:= -E`1[`f^{(0)}_{\RX,\RY}(`q)`2]+\ln E`1[e^{`f^{(0)}_{\RX',\RY'}}`2]\\
  &\kern1em-E`1[`f^{(1)}_{\RX}(`q)`2]+\ln E`1[e^{`f^{(1)}_{\RX'}}`2]\\
  &\kern1em-E`1[`f^{(2)}_{\RY}(`q)`2]+\ln E`1[e^{`f^{(2)}_{\RY'}}`2].
\end{align*}

\paragraph{Mutual information estimate}
The mutual information can then be estimated with
\begin{subequations}
  \label{eq:est:I}
  \begin{align}
  I(\RX\wedge \RY) &\approx \frac1N \sum_{i=1}^N `f^{(0)}_{\RX_i,\RY_i}(`q_t) - \ln \frac1{N'} \sum_{i=1}^{N'} e^{`f^{(0)}_{\RX'_i,\RY'_i}(`q_t)}\label{eq:est:I1}\\
  & \kern0em- \frac1N \sum_{i=1}^N `f^{(1)}_{\RX_i}(`q_t) + \ln \frac1{N'} \sum_{i=1}^{N'} e^{`f^{(1)}_{\RX'_i}(`q_t)}\label{eq:est:I2}\\
  & \kern0em- \frac1N \sum_{i=1}^N `f^{(2)}_{\RY_i}(`q_t) + \ln \frac1{N'} \sum_{i=1}^{N'} e^{`f^{(2)}_{\RY'_i}(`q_t)},\label{eq:est:I3}
  \end{align}
\end{subequations}
where $`q_t$ is the parameter after training the neural network $t$ times.

\subsection{Estimation using a uniform reference}
\label{subsec:unif}

MINE can be viewed as the special case of the mutual information estimation in the last section when the reference distribution is chosen as the product of marginal distribution of $\RX$ and $\RY$, i.e., 
\begin{align}
  p_{\RX'\RY'}(x,y)=p_{\RX}(x)p_{\RY}(y) \quad \forall x\in X, y\in Y.\label{eq:MINE}
\end{align}
In this case, both $D(p_{\RX}\|p_{\RX'})$ and $D(p_{\RY}\|p_{\RY'})$ in \eqref{eq:I3} are zero, and so
\begin{subequations}
  \label{eq:I34:MINE}
  \begin{align}
    I(\RX\wedge \RY) &= D(p_{\RX\RY}\| p_{\RX'\RY'})
    \label{eq:I3:MINE}\\
    \begin{split}&= 
      \sup_{f_0:X\times Y\to `R} \{E`1[f_0(\RX,\RY)`2] - \ln E`1[ e^{f_0(\RX',\RY')}`2]\},
    \end{split}\label{eq:I4:MINE}
  \end{align}
\end{subequations}
and the optimal solution $f_0$ satisfies
\begin{align}
  f_0(x,y) = i_{\RX\RY}(x,y) + c \label{eq:f*:I:MINE}
\end{align}
for some constant $c$, where $i_{\RX\RY}(x,y)$ is defined in~\eqref{eq:f*:I}. With the optimal $f_0$, the first term in \eqref{eq:I4:MINE} becomes $E`1[f_0(\RX,\RY)`2]=I(\RX\wedge \RY) + c$, namely a constant shift of the mutual information, while the second term becomes $-\ln E`1[ e^{f_0(\RX',\RY')}`2]=-c$, which cancels out the constant shift to give the desired mutual information. 

Note that there is no need to train the neural network for the outputs $`f^{(1)}_x$ and $`f^{(2)}_y$ because the corresponding terms~\eqref{eq:est:I2} and \eqref{eq:est:I3} do not appear in \eqref{eq:I4:MINE}. To train the remaining output $`f^{(0)}_{x,y}$, one cannot sample $(\RX',\RY')$ from the unknown pdf's\ $p_{\RX}$ and $p_{\RY}$.
Instead, as done in MINE, the samples $\RX'_i$'s and $\RY'_i$'s can be obtained by resampling the samples $\RX_i$'s and $\RY_i$'s independently. As a result, one cannot arbitrarily reduce the bias in estimating the log expectation term and its gradient in \eqref{eq:I4:MINE}. Different from MINE, we choose the following uniform distribution.

\paragraph{Uniform reference} We obtain the mutual information estimate with 
\begin{subequations}
  \label{eq:uniform} 
  \begin{align}
    p_{\RX'\RY'}(x,y) = p_{\RX'}(x)p_{\RY'}(y) 
    &= \begin{cases}
      \frac1{\op{Vol}(\rmB)} & (x,y) \in \rmB\\
      0 & \text{otherwise},
    \end{cases}\label{eq:uniform:pdf}
  \end{align}
  where $B$ is a bounding box with volume $\op{Vol}(\rmB)$ and containing all the values of $(x,y)$ with 
  \begin{align}
    \begin{split}
    &\min_{1\leq i\leq N} \RX_i\leq x \leq \max_{1\leq i\leq N} \RX_i\\
    &\min_{1\leq i\leq N} \RY_i\leq y \leq \max_{1\leq i\leq N} \RY_i.
    \end{split}\label{eq:B}
  \end{align}
\end{subequations}
If $\RX$ and $\RY$ are vectors, the above minimization, maximization, and inequalities are elementwise. 

There is, however, a technical issue with the above choice of uniform reference. \eqref{eq:I4} is valid only if \eqref{eq:supp:I} holds, which requires $\rmB$ to contain all $(x,y)$ with $p_{\RX\RY}(x,y)>0$. However, such requirement may not be satisfied as $p_{\RX\RY}$ is unknown and may have unbounded support. Nevertheless, we argue that the above choice of $\rmB$ can still give a good estimate if the density $p_{\RX\RY}$ outside $\rmB$ has negligible contribution to the mutual information. More precisely, define $(\tRX,\tRY)$ with density
\begin{align*}
  p_{\tRX\tRY}(x,y)=\frac{p_{\RX\RY}(x,y)}{\Pr\Set{(\RX,\RY)\in \rmB}} \quad \forall (x,y)\in \rmB,
\end{align*}
namely the conditional density of $(\RX,\RY)$ given $(\RX,\RY)\in \rmB$. Note that $\Pr\Set{(\RX,\RY)\in \rmB}$ goes to $1$ as $N$ goes to infinity by \eqref{eq:B}. 
We therefore make the mild assumption that
\begin{align}
I(\tRX\wedge \tRY) \approx I(\RX\wedge \RY)\label{eq:uniform:assumption}
\end{align}
for sufficiently large $N$.
Since $(\RX_i,\RY_i)$ can also be viewed as samples of $(\tRX,\tRY)$, we can estimate $I(\tRX\wedge \tRY)$ using the same formula~\eqref{eq:est:I}. In particular, it is valid to use a uniform reference because its support covers that of $p_{\tRX,\tRY}$.

\section{Experimental results}
\label{sec:experiment}

To evaluate the convergence rate, we plotted the mutual information estimates~\eqref{eq:est:I} with uniform reference~\eqref{eq:uniform} against the number of training steps and compared the curve to that of MINE. We first consider a simple bivariate mixed gaussian distribution and show that MINE has much slower convergence than our approach even in this low dimensional example. We then consider the higher dimensional case using a basic gaussian distribution and show that our approach can achieve significantly faster convergence rate even with a moderate increase in the dimension. 

% To evaluate whether the neural network converges to the desired optimal solution, we also plotted
% \begin{align}
%   \hat{i}_{`q_t}(x,y):=`f^{(0)}_{x,y}(`q_t) - `f^{(1)}_x(`q_t) - `f^{(2)}_y(`q_t),\label{eq:est:i}
% \end{align}
% which should ideally converge to the R.H.S.\ of $i_{\RX\RY}(x,y)+c$~in \eqref{eq:f*:I} for some constant $c$.
% For bivariate distributions, we can plot the function for different $(x,y)\in \rmB$~\eqref{eq:B} as a heatmap for each step $t$. We can then relate the convergence rate at different steps to the learning of different regions of $i_{\RX\RY}(x,y)$.

The bivariate mixed gaussian distribution is defined as
\begin{align}
  \op{MG}(`r):\quad p_{\RX\RY}(x,y) = \frac12 \mcN_{\bSM x\\ y\eSM }`1(\M0,\bM 1 & `r\\ `r & 1\eM`2) + \frac12 \mcN_{\bSM x\\ y\eSM}`1(\M0,\bM 1 & -`r\\ -`r & 1\eM`2)\label{eq:mg}
\end{align}
where $\mcN_{\Mz}(\M\mu,\M{\Sigma})$ denotes the multivariate gaussian distribution over $\Mz$ with mean $\M\mu$ and covariance matrix $\M{\Sigma}$, and $`r\in [0,1)$ is a model parameter that specifies the positive and negative correlations of $\RX$ and $\RY$ for each gaussian component. The higher dimensional gaussian distribution is defined as
\begin{align}
  \op{HG}(`r,d):\quad p_{\RX\RY}(\Mx,\My) = \prod_{i=1}^d \mcN_{\bSM x_i\\ y_i\eSM }`1(\M0,\bM 1 & `r\\ `r & 1\eM`2) \quad \text{where}\kern1em \begin{aligned}\Mx&:=(x_1,\dots,x_d),\\
  \My&:=(y_1,\dots,y_d)\in `R^d.\end{aligned}\label{eq:hg}
\end{align}
In addition to the correlation coefficient $\rho$, there is an additional parameter $d$ that specifies the dimension of $\RX$ and $\RY$. 

\begin{figure}[h!]
  \subcaptionbox{MI-NEE.\label{fig:mg:rho=0.9:mi:MINEE}}[.5\linewidth]{\includegraphics[width=6.9cm]{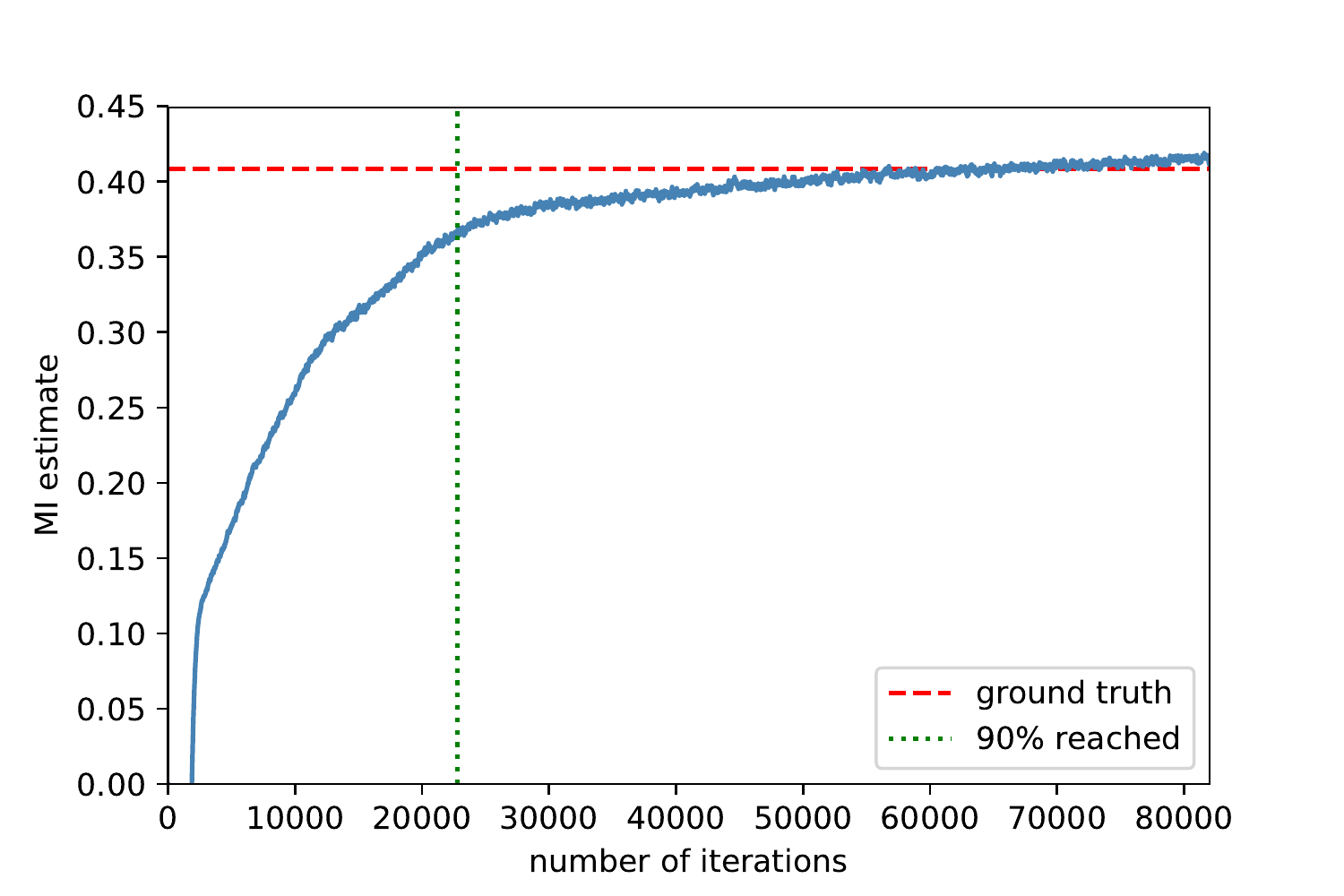}}
  \subcaptionbox{MINE.\label{fig:mg:rho=0.9:mi:MINE}}[.5\linewidth]{\includegraphics[width=6.9cm]{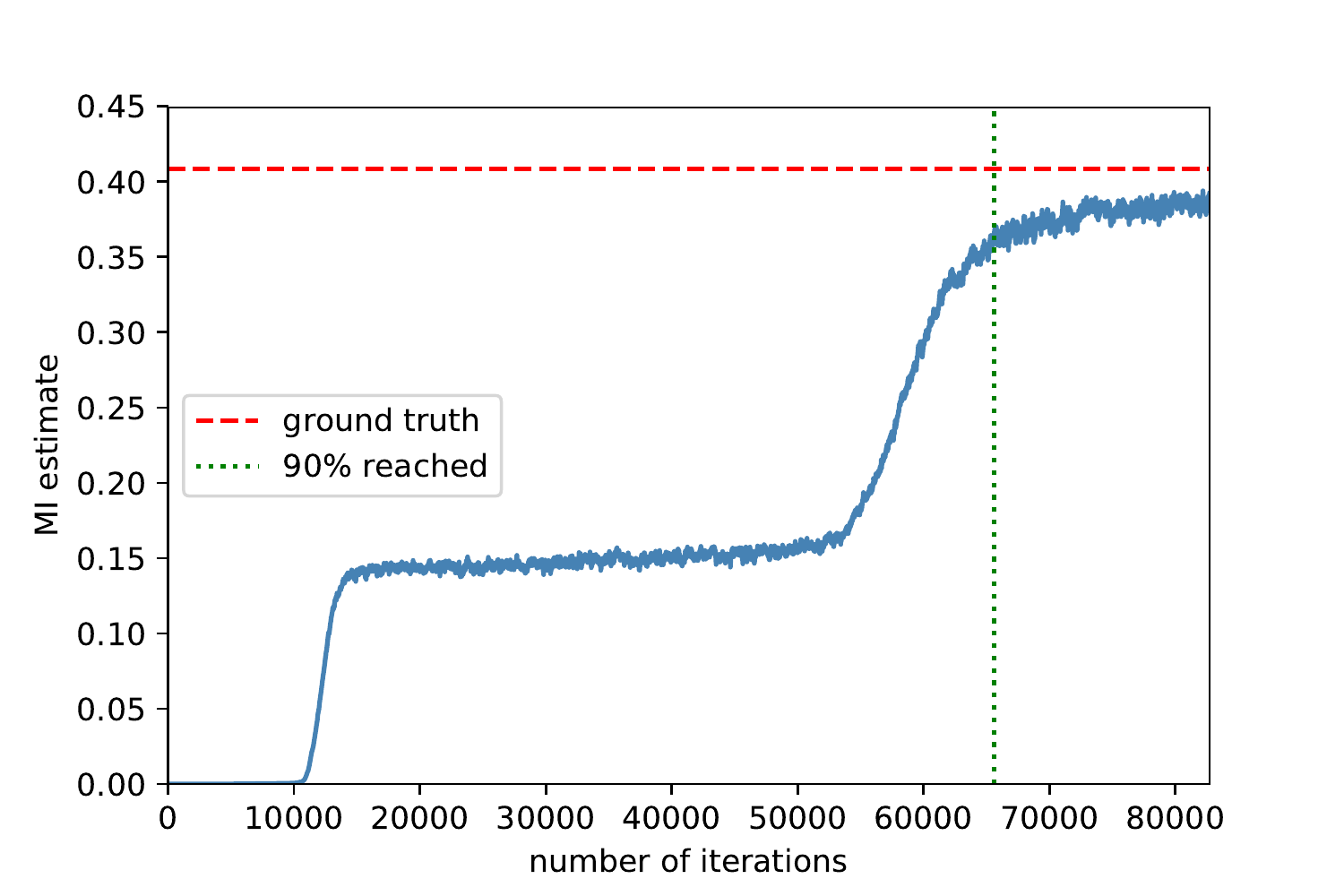}}
  \caption{The mutual information estimates for $\op{MG}(0.9)$.}
  \label{fig:mg:rho=0.9:mi}
\end{figure}

For the mixed gaussian model $\op{MG}(0.9)$ with sample size $N=400$ points, Figure~\ref{fig:mg:rho=0.9:mi} plots the mutual information estimates after training with a batch size of $100$ and learning rate of $10^{-4}$. For MINE, we follow \cite{belghazi18} to use moving average in the gradient estimate, where the moving average rate is set to be $0.01$. For our approach, instead of using a moving average in the gradient estimate, we increase the reference sample size $N'$ to $10$ times the data sample size $N$. For both MINE and our approach, we further apply a moving average of rate $0.01$ to smooth out jitters in the estimates. Figure~\ref{fig:mg:rho=0.9:mi:MINEE} shows that our approach converges to within $10\%$ of the ground truth close to $2\times 10^4$ iterations. Figure~\ref{fig:mg:rho=0.9:mi:MINE} shows that MINE requires close to $7\times 10^4$ iterations. Furthermore, MINE exhibits a staircase convergence with two distinct jumps. The estimate remains close to $0$ until the first jump at around $10^4$ iterations. The estimate then remains stagnant at a value smaller than $50\%$ of the ground truth until the second jump at around $5\times 10^4$ iterations. We remark that the staircase convergence may mislead one to think that neural network has converged while it has not. We found that the issue can be more serious for smaller values of $\rho$. 

\begin{figure}[h!]
  \subcaptionbox{MI-NEE.\label{fig:hg:rho=0.9:mi:MINEE}}[0.5\linewidth]{\includegraphics[width=6.9cm]{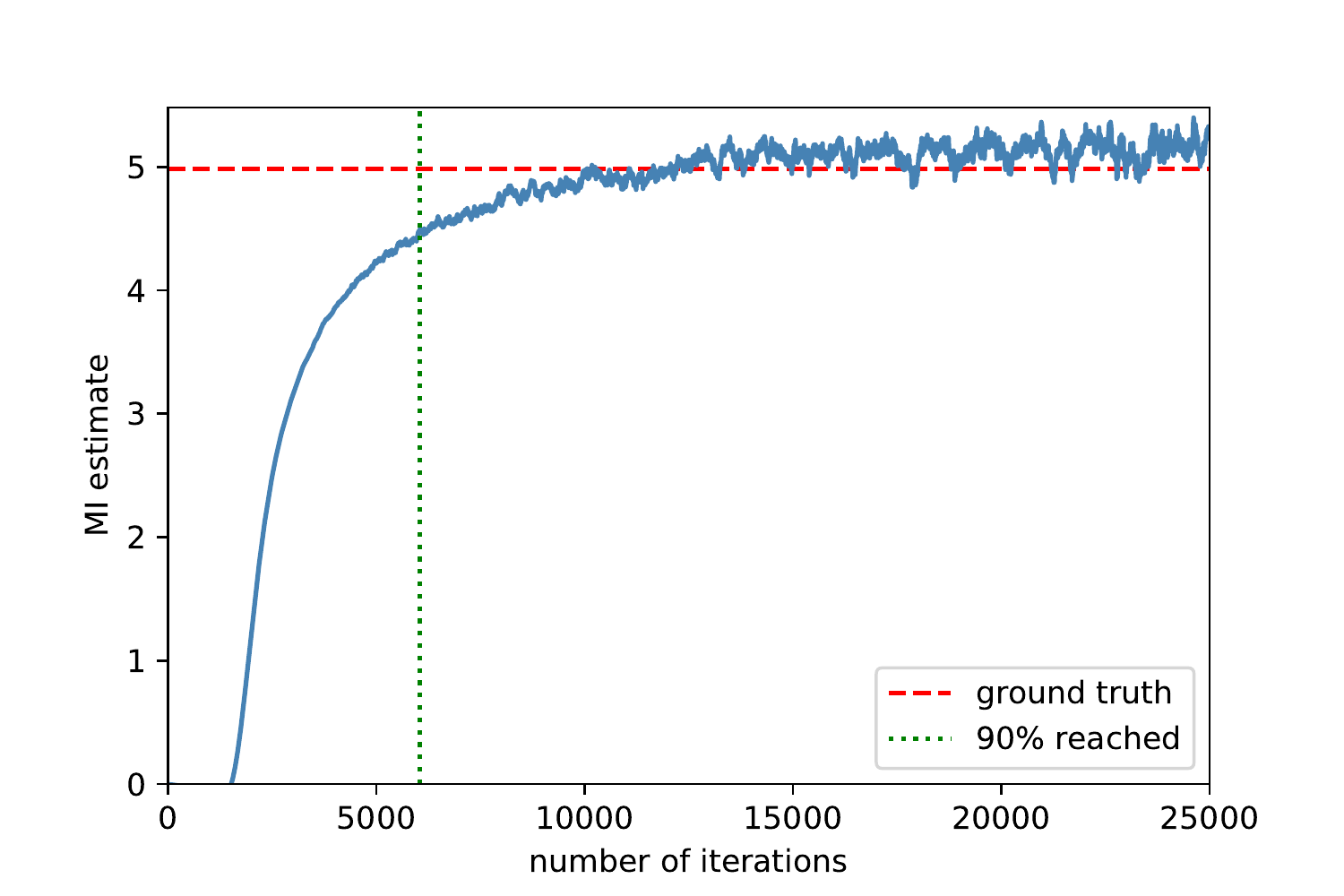}}
  \subcaptionbox{MINE.\label{fig:hg:rho=0.9:mi:MINE}}[0.5\linewidth]{\includegraphics[width=6.9cm]{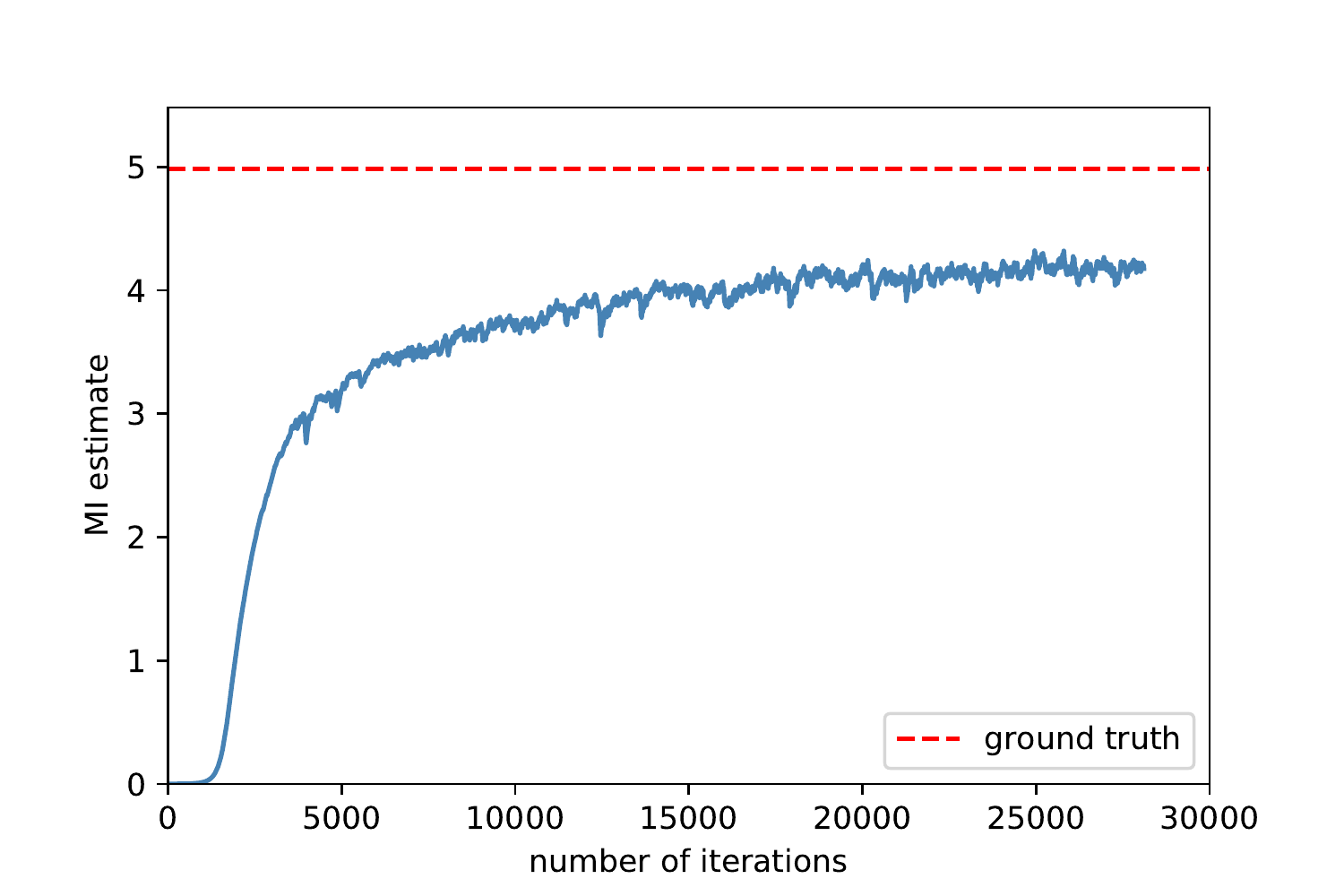}}
  \caption{The mutual information estimates for $\op{HG}(0.9,6)$.}
  \label{fig:hg:rho=0.9:mi}
\end{figure}

For the higher dimensional gaussian distribution, we consider $\op{HG}(0.9,6)$ with again a sample size of $400$ and a batch size of $100$. The learning rate is reduced to $5\times 10^{-5}$ to avoid excessive jitters. For our approach, we increase the reference sample size to $300$ times the data sample size to reduce the effect of overfitting the reference. Figure~\ref{fig:hg:rho=0.9:mi:MINEE} shows that our approach converges to within $10\%$ of the ground truth close to $6\times 10^3$ iterations. However, Figure~\ref{fig:hg:rho=0.9:mi:MINE} shows that MINE is unable to converge to within $10\%$ of the ground truth even after $2.5\times 10^4$ iterations.\footnote{In \cite[Fig.~1]{belghazi18}, MINE reaches about $20\%$ of the ground truth, however, we were unable to reproduce this results since, to the best of our knowledge, the authors' parameters choice / code are not publicly available. Nevertheless, our observations remain valid since the comparisons made here between MINE and MI-NEE are performed under comparable parameters / neural network architecture.} Indeed, MINE terminates before $3\times 10^4$ iterations due to numerical instability issue, but further reducing the learning rate causes excessive slow down in convergence. In contrast, our approach has slight overfitting as the estimate can go above the ground truth. We found that this issue is more pronounced for higher dimension, but can be alleviated by increasing the reference sample size in the expense of more computations for each training step. One can also use a separate validation set to terminate the training of each neural networks before significant overfitting.

The above results can be reproduced by running the corresponding jupyter notebooks using binder \cite{jupyter2018binder} at the GitHub repository below:
\begin{center}
\url{https://github.com/ccha23/MI-NEE}
\end{center}

\clearpage
\bibliographystyle{abbrv}
\bibliography{IEEEabrv,ref}

% \appendix

% \section{Anonymous Github repository}
% \url{https://github.com/fasterminee/NIPS2019}
% \section{Other experimental results}
% video?

% Include videos if available, but need anonymous links.
% If needed, can use separate files in supplementary for Other experiment results. 

\end{document}